\documentclass[journal=cmatex,manuscript=article]{achemso}

\usepackage{achemso}
\usepackage[version=3]{mhchem} 
\usepackage{amssymb}
\usepackage{color}
\author{Zhi-Cheng Wang}
\affiliation{Department of Physics, Zhejiang University, Hangzhou 310027, China}
\author{Chao-Yang He}
\affiliation{Department of Physics, Zhejiang University, Hangzhou 310027, China}
\author{Si-Qi Wu}
\affiliation{Department of Physics, Zhejiang University, Hangzhou 310027, China}
\author{Zhang-Tu Tang}
\affiliation{Department of Physics, Zhejiang University, Hangzhou 310027, China}
\author{Yi Liu}
\affiliation{Department of Physics, Zhejiang University, Hangzhou 310027, China}
\author{Guang-Han Cao}
\affiliation{Department of Physics, Zhejiang University, Hangzhou 310027, China}
\alsoaffiliation{State Key Lab of Silicon Materials, Zhejiang University, Hangzhou 310027, China}
\alsoaffiliation{Collaborative Innovation Centre of Advanced Microstructures, Nanjing 210093, China}
\email{ghcao@zju.edu.cn}

\title[]{Synthesis, crystal structure and superconductivity in Rb$Ln_2$Fe$_4$As$_4$O$_2$ ($Ln$ = Sm, Tb, Dy and Ho)}

\begin{document}








\begin{abstract}
We have synthesized four iron-based oxyarsenide superconductors Rb$Ln_2$Fe$_4$As$_4$O$_2$ ($Ln$ = Sm, Tb, Dy and Ho) resulting from the intergrowth of RbFe$_2$As$_2$ and $Ln$FeAsO. It is found that the lattice match between RbFe$_2$As$_2$ and $Ln$FeAsO is crucial for the phase formation. The structural intergrowth leads to double asymmetric Fe$_2$As$_2$ layers that are separated by insulating $Ln_2$O$_2$ slabs. Consequently, the materials are intrinsically doped at a level of 0.25 holes/Fe-atom and, bulk superconductivity emerges at $T_\mathrm{c}$ = 35.8, 34.7, 34.3 and 33.8 K, respectively, for $Ln$ = Sm, Tb, Dy and Ho. Investigation on the correlation between crystal structure and $T_\mathrm{c}$ suggests that interlayer couplings may play an additional role for optimization of superconductivity.
\end{abstract}

\newpage
\section{INTRODUCTION}

Recent years have witnessed discoveries of many iron-based superconductors (IBS) crystallizing in several structure types\cite{hosono-pc,cxh-scm,jh}. The key structural unit for the emergence of superconductivity is the anti-fluorite-type Fe$_2$$X_2$ ($X$ = As, Se) layers, with which the parent (undoped) compounds mostly appear to be spin-density-wave (SDW) semi-metals. Superconductivity is induced by suppressing the SDW ordering via a certain chemical doping that may either introduce additional electrons\cite{hosono} or holes\cite{whh}, or ``apply" chemical pressures\cite{rz}. Nevertheless, there is an alternative route towards superconductivity as well, namely, the electron (or hole) carriers are introduced by an internal charge transfer in the material itself. Examples include self-electron-doped Sr$_2$VFeAsO$_3$\cite{cgh} and Ba$_2$Ti$_2$Fe$_2$As$_4$O\cite{syl}. More recent examples are manifested by the 1144-type $AkAe$Fe$_4$As$_4$ ($Ak$ = K, Rb, Cs; $Ae$ = Ca, Sr, Eu)\cite{1144,Eu1144,RbEu1144,CsEu1144} and 12442-type KCa$_2$Fe$_4$As$_4$F$_2$\cite{wzc}, both of which are self-hole-doped owing to charge homogenization.

\begin{figure}
\center
\includegraphics[width=10cm]{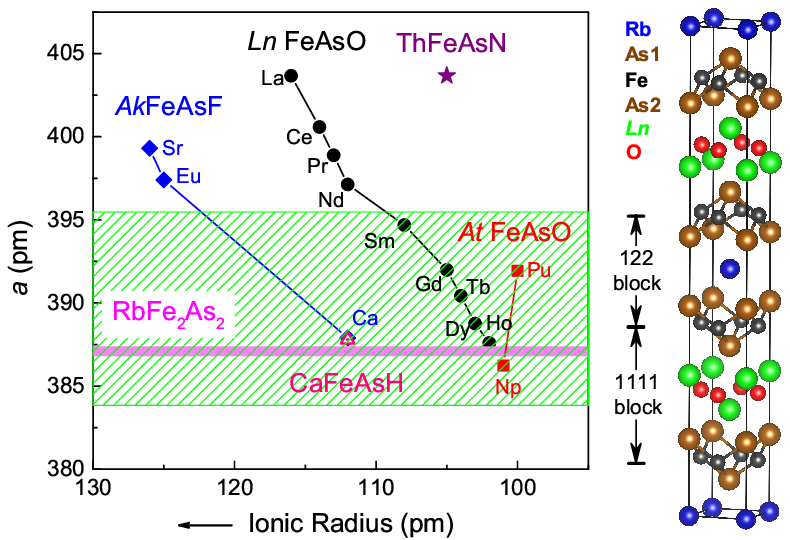}
\caption{\label{match} Lattice matching between 122-type RbFe$_2$As$_2$ and 1111-type $Ln$FeAsO ($Ln$ = lanthanide elements, data taken from Ref.~\cite{LnO}). Based on experimental results, the range for formation of Rb$Ln_2$Fe$_4$As$_4$O$_2$ is marked by the shaded area, where a good lattice match of  RbFe$_2$As$_2$ and $Ln$FeAsO is satisfied. The $a$ parameters of CaFeAsH\cite{CaH}, $Ae$FeAsF ($Ae$ = Ca, Sr and Eu)\cite{CaF,SrF}, $At$FeAsO ($At$ = Np and Pu)\cite{NpO,PuO} and ThFeAsN\cite{wc2016} are put together for comparison. The horizontal axis denotes the ionic radii of $Ae^{2+}$ [coordination number ($CN$) = 8], $Ln^{3+}$ ($CN$ = 8) and $At^{3+}$ ($CN$ = 6)\cite{shannon}. Shown at the right side is the 12442-type structure composed of 122- and 1111-blocks.}
\end{figure}

We previously formulated a strategy of structural design for the exploration of new IBS\cite{jh}, which helps to discover the KCa$_2$Fe$_4$As$_4$F$_2$ superconductor\cite{wzc}. KCa$_2$Fe$_4$As$_4$F$_2$ can be viewed as an intergrowth of 1111-type CaFeAsF and 122-type KFe$_2$As$_2$, as shown on the right side of Fig.~\ref{match}. The resulting crystal structure possesses double Fe$_2$As$_2$ layers that are separated by the insulating fluorite-type Ca$_2$F$_2$ slab, mimicking the case of double CuO$_2$ planes in cuprate superconductors. Note that the CaFeAsF slab is undoped, while the KFe$_2$As$_2$ block is heavily hole doped (0.5 holes/Fe-atom). As a result, the hybrid structure intrinsically bears hole doping at 25\%, which leads to absence of SDW ordering and appearance of superconductivity at 33 K\cite{wzc}. Along this research line, we succeeded in synthesizing two additional quinary fluo-arsenides $Ak$Ca$_2$Fe$_4$As$_4$F$_2$ ($Ak$ = Rb and Cs) with $T_\mathrm{c}$ = 30.5 K and 28.2 K, respectively\cite{wzc-scm}. Furthermore, we also obtained the first 12442-type oxyarsenide RbGd$_2$Fe$_4$As$_4$O$_2$ which superconducts at 35 K\cite{wzc-Gd}. Questions arise naturally: can 12442-type oxyarsenides be synthesized if Gd is replaced by any other lanthanide elements? How does $T_\mathrm{c}$ change with such element replacements? Whether or not the lanthanide magnetism influences the $T_\mathrm{c}$?

As we previously pointed out, lattice match between the constituent crystallographic block layers is crucial to realize the designed structures\cite{jh}. So, let us first investigate the lattice-match issue. Fig.~\ref{match} plots lattice parameter $a$ of various 1111-type Fe$_2$As$_2$-layer containing compounds, in comparison with that of RbFe$_2$As$_2$. The horizontal axis shows the effective ionic radii of $Ae^{2+}$, $Ln^{3+}$, $At^{3+}$ ($At$ = Np and Pu) and Th$^{4+}$\cite{shannon}. The data clearly explain why only Ca-containing fluo-arsenides $Ak$Ca$_2$Fe$_4$As$_4$F$_2$ were obtained (syntheses of $AkAe_2$Fe$_4$As$_4$F$_2$ with $Ae$ = Sr or Eu at ambient pressure were unsuccessful). Given the formation of RbGd$_2$Fe$_4$As$_4$O$_2$ at ambient pressure\cite{wzc-Gd}, one expects from the plot that synthesis of Rb$Ln_2$Fe$_4$As$_4$O$_2$ with $Ln$ = Tb and Dy is very likely (because of better lattice match). We actually succeeded in synthesizing four new members of Rb$Ln_2$Fe$_4$As$_4$O$_2$ with $Ln$ = Sm, Tb, Dy and Ho, among which RbHo$_2$Fe$_4$As$_4$O$_2$ is notable because the constituent HoFeAsO cannot be prepared at ambient pressure. In this paper, we report synthesis, crystal structure and superconductivity of the four new materials. Influences of crystal structure and lanthanide magnetism on $T_\mathrm{c}$ are discussed.

\section{EXPERIMENTAL SECTION}

We attempted to synthesize seven target compounds Rb$Ln_2$Fe$_4$As$_4$O$_2$ with $Ln$ = Nd, Sm, Tb, Dy, Ho, Er and Y, by employing a solid-state reaction method, similar to our previous report\cite{wzc}. The source materials include Rb ingot (99.75\%), $Ln$ ingot (99.9\%), $Ln_2$O$_3$ and Tb$_4$O$_7$ powders (99.9\%), Fe powders (99.998\%) and As pieces (99.999\%). Intermediate products of $Ln$As, FeAs and Fe$_2$As were presynthesized by direct solid-state reactions of their constituent elements. RbFe$_2$As$_2$ was additionally prepared by reacting Rb ingot (with an excess of 3\%) and FeAs at 923 K for 10 hours. With these intermediate products, Rb$Ln_2$Fe$_4$As$_4$O$_2$ samples were finally synthesized by solid-state reactions of the stoichiometric mixtures of RbFe$_2$As$_2$, $Ln$As, $Ln_2$O$_3$, Tb$_4$O$_7$, FeAs and Fe$_2$As. The chemical reactions take place in a small alumina container which is sealed in a Ta tube. The Ta tube was further jacketed with a quartz ampoule. This sample-loaded ampoule was sintered at 1213 - 1253 K for 40 hours, after which it was allowed to cool down by switching off the furnace.

Powder x-ray diffraction (XRD) was carried out on a PANAlytical x-ray diffactometer with a CuK$_{\alpha1}$ monochromator at room temperature. To obtain the crystallographic data of the new compounds Rb$Ln_2$Fe$_4$As$_4$O$_2$ with $Ln$ = Nd, Sm, Tb, Dy and Ho, we made a Rietveld refinement employing the software RIETAN-FP\cite{rietan}. The 12442-type structural model\cite{wzc} was adopted to fit the XRD data in the range of 20$^{\circ}\leq 2\theta \leq 150^{\circ}$. The occupation factor of each atom was fixed to 1.0. As a result, the converged refinement yields fairly good reliable factors of $R_\mathrm{wp}$ = 2.96\% ($Ln$ = Sm), 2.98\% ($Ln$ = Tb), 2.46\% ($Ln$ = Dy) and  2.60\% ($Ln$ = Ho), and goodness-of-fit parameters of $S$ = 1.18 ($Ln$ = Sm), 1.01 ($Ln$ = Tb), 1.14 ($Ln$ = Dy) and  1.03 ($Ln$ = Ho).

We employed a physical property measurement system (Quantum Design, PPMS-9) and a magnetic property measurement system (Quantum Design, MPMS-XL5) for the measurements of temperature dependence of electrical resistance and magnetic moments. A standard four-electrode method and the ac transport option were utilized for the resistivity measurement. Samples for the magnetic measurements were cut into regular shape so that the demagnetization factors can be accurately estimated. To detect superconducting transitions, we applied a low field of 10 Oe in both zero-field cooling (ZFC) and field cooling (FC) modes. The isothermal magnetization curves above and well below $T_\mathrm{c}$ were measured. We also measured the temperature dependence of magnetic susceptibility up to room temperature under an applied field of 5000 Oe.

\section{RESULTS AND DISCUSSION}

Our XRD experiments indicate that the expected 12442-type Rb$Ln_2$Fe$_4$As$_4$O$_2$ can be successfully synthesized at ambient pressure for $Ln$ = Sm, Tb, Dy and Ho. In the case of $Ln$ = Nd, however, only RbFe$_2$As$_2$ and NdFeAsO show up in the final product. This fact suggests that the lattice mismatch between RbFe$_2$As$_2$ and NdFeAsO, as shown in Fig.~\ref{match}, is so heavy that RbNd$_2$Fe$_4$As$_4$O$_2$ is no longer stable at the ambient-pressure synthesis condition. From this empirical result, the criterion for possible formation of 12442-type phases is that the lattice mismatch, defined as $2(a_{1111}-a_{122})/(a_{1111}+a_{122})$, is less than 2\%. For $Ln$ = Er (Y), the resulting phases are RbFe$_2$As$_2$, Er$_2$O$_3$ (Y$_2$O$_3$), ErAs (YAs), Fe$_2$As and FeAs, although good lattice match between RbFe$_2$As$_2$ and ``ErFeAsO" is expected (from extrapolation). The failure of synthesis of RbEr$_2$Fe$_4$As$_4$O$_2$ (RbY$_2$Fe$_4$As$_4$O$_2$) is then mainly due to the instability of the ErFeAsO (YFeAsO) block. For this reason, the successful synthesis of RbHo$_2$Fe$_4$As$_4$O$_2$ is remarkable, because HoFeAsO by itself cannot be synthesized at ambient pressure. Interestingly we sometimes observe HoFeAsO as a secondary phase when synthesizing RbHo$_2$Fe$_4$As$_4$O$_2$. This HoFeAsO phase could form as a result of decomposition of RbHo$_2$Fe$_4$As$_4$O$_2$ during the high-temperature solid-state reactions.

Given that NpFeAsO\cite{NpO} and PuFeAsO\cite{PuO} can be synthesized at ambient pressure and, their lattices well match that of RbFe$_2$As$_2$ (see Fig.~\ref{match}), syntheses of Rb$At_2$Fe$_4$As$_4$O$_2$ ($At$ = Np and Pu) are very likely. It is of particular interest whether these 12442 species superconduct or not, since superconductivity is absent in the actinide-containing 1111 systems\cite{Np1111}. Noted also is the lattice match for CaFeAsH\cite{CaH} (though it was synthesized at high pressures), as such, RbCa$_2$Fe$_4$As$_4$H$_2$ is also expectable. By employing high-pressure synthesis technique, additional 12442 members such as RbNd$_2$Fe$_4$As$_4$O$_2$ and RbY$_2$Fe$_4$As$_4$O$_2$ might also be synthesized. Furthermore, one may extend the synthesis to K- and Cs-containing 12442 series, similarly by the consideration of lattice match between KFe$_2$As$_2$ (or CsFe$_2$As$_2$) and $Ln$FeAsO. Such studies are under going.

\begin{figure}
\center
\includegraphics[width=10cm]{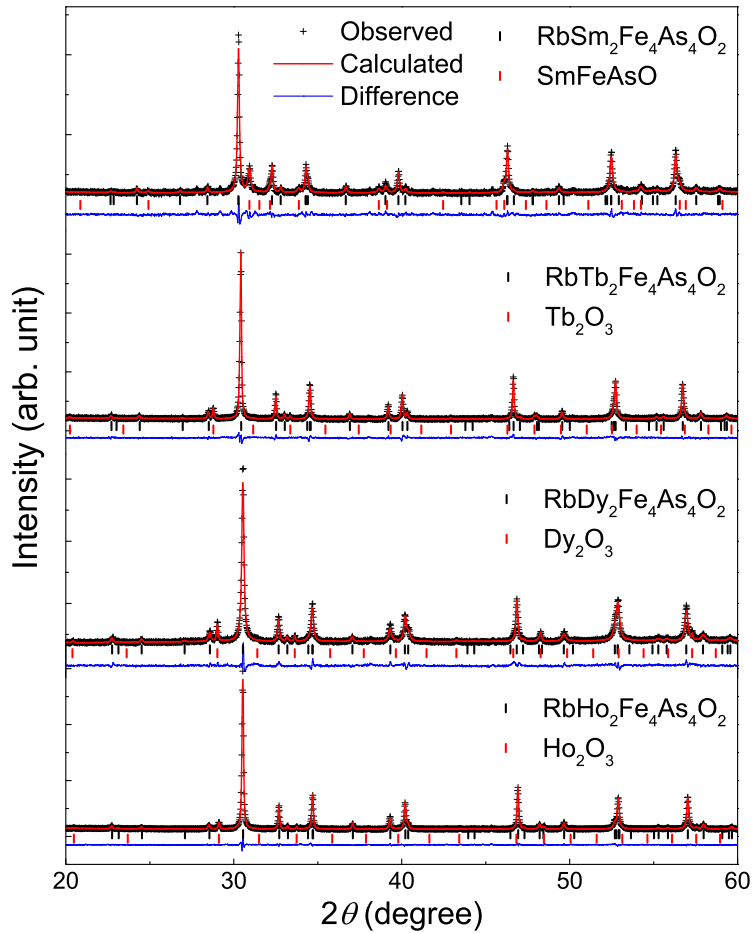}
\caption{\label{xrd} Powder X-ray diffraction patterns and their Rietveld refinement profiles for Rb$Ln_2$Fe$_4$As$_4$O$_2$ ($Ln$ = Sm, Tb, Dy and Ho). Only low-angle ($20^{\circ}\leq 2\theta \leq 60^{\circ}$) data are shown to highlight the main reflections.}
\end{figure}

Figure~\ref{xrd} shows the XRD patterns of Rb$Ln_2$Fe$_4$As$_4$O$_2$ with $Ln$ = Sm, Tb, Dy and Ho. Most of the reflections can be indexed with a body-centered tetragonal lattice of $a \approx$ 3.90 {\AA} and $c \approx$ 31.3 {\AA}, consistent with the 12442-type structure\cite{wzc}. Samples of $Ln$ = Tb, Dy and Ho are nearly single phase. The detectable impurity is $Ln_2$O$_3$ whose weight percentages are 3.6 \%, 3.9 \% and 2.8 \%, respectively, according to our Rietveld analyses. For $Ln$ = Sm, synthesis of high-purity sample is difficult (the proportion of the main secondary phase SmFeAsO is 16.4 wt.\% for the sample reported here). This might reflect that RbSm$_2$Fe$_4$As$_4$O$_2$ locates at the verge of chemical instability. The structural refinement results are listed in Table~\ref{structure} where the data of RbGd$_2$Fe$_4$As$_4$O$_2$\cite{wzc-Gd} is also included for comparison.

\begin{table*}
\caption{\label{structure}
Room-temperature crystallographic data of Rb$Ln_2$Fe$_4$As$_4$O$_2$ in comparison with each other. The space group is $I4/mmm$ (No. 139). The atomic coordinates are as follows: Rb 2$a$(0, 0, 0); \textit{Ln} 4$e$(0.5, 0.5, $z$); Fe 8$g$ (0.5, 0, $z$); As1 4$e$(0.5, 0.5, $z$); As2 4$e$(0, 0, $z$);  O 4$d$(0.5, 0, 0.25).}
\center
\begin{tabular}{lccccc}
\hline\hline
$Ln$ & Sm & Gd (Ref.~\cite{wzc-Gd}) & Tb & Dy & Ho\\
\hline
Lattice Parameters&  & & & & \\\hline
$a$ (\r{A})  &3.9209(2) & 3.9014(2) &3.8900(1) & 3.8785(2) & 3.8688(1)\\
$c$ (\r{A})  &31.381(2) & 31.343(2) &31.277(1) & 31.265(2) & 31.2424(7) \\
$V$ (\r{A}$^{3}$)   & 482.44(5)& 477.06(4)  &473.29(3) &470.30(4)& 467.64(2) \\
\hline
Coordinates ($z$)& &  &  & \\ \hline
\textit{Ln} & 0.2127(1) &0.2138(1)  &  0.21394(8) & 0.21382(9)& 0.21414(6) \\
      Fe &0.1131(2) & 0.1138(2) &0.11461(12)&0.11525(12) &0.11570(7)\\
      As1 &0.0707(2)& 0.0697(2) &0.06990(14)&0.06907(17)&0.06961(8)\\
      As2  &0.1575(2)& 0.1591(2) &0.15990(11) &0.15983(13)&0.16051(7) \\
\hline
Bond Distances&& &&  \\\hline
Fe$-$As1 (\r{A}) & 2.369(5) &  2.391(6) &2.396(3)  &2.418(5)& 2.412(2) \\
Fe$-$As2 (\r{A}) & 2.405(6) & 2.413(6)  & 2.406(3) &2.388(3)& 2.388(3) \\
\hline
As Height&& &&  \\\hline
As1 (\r{A}) &1.331(13) &  1.382(13)  & 1.398(7) &1.444(10)&1.440(5)  \\
As2 (\r{A}) &1.387(13)  & 1.420(13) &1.417(7) & 1.394(7)&1.400(7)\\
\hline
Bond Angles && &&  \\\hline
As1$-$Fe$-$As1 ($^{\circ}$) & 111.7(3) & 109.4(3)& 108.6(2) &106.6(2)&106.7(1)  \\
As2$-$Fe$-$As2 ($^{\circ}$) &109.2(3) &  107.9(3)&107.9(2) &108.7(2)&108.2(1)   \\
\hline
Fe$_2$As$_2$-Layer Spacings&& &&  \\\hline
$d_{\mathrm{intra}}$ (\r{A}) &7.098(9)&7.134(10)&7.169(4)&7.207(5)&7.229(5)\\
$d_{\mathrm{inter}}$ (\r{A})&8.592(9)&8.538(10)&8.469(4)&8.426(5)&8.392(5)\\
\hline\hline
\end{tabular}
\end{table*}

\begin{figure}
\center
\includegraphics[width=10cm]{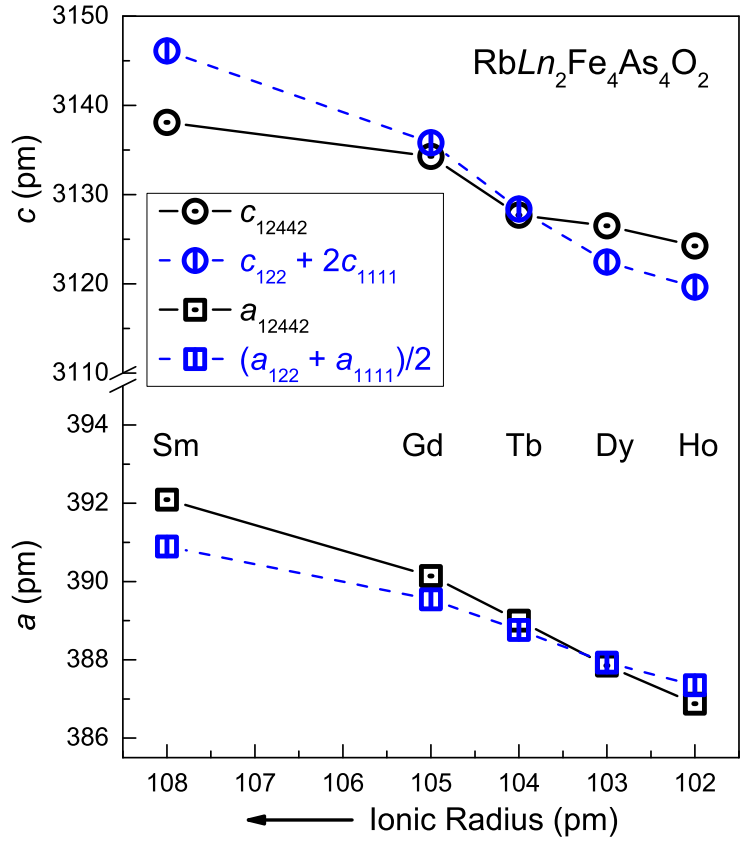}
\caption{\label{lattice} Lattice parameters of Rb$Ln_2$Fe$_4$As$_4$O$_2$ ($Ln$ = Sm, Gd, Tb, Dy and Ho) as a function of ionic radii of $Ln^{3+}$. The symbols in blue with dashed lines denote the average values of their constituent 122-type and 1111-type unit cells.}
\end{figure}

Fig.~\ref{lattice} shows lattice parameters $a$ and $c$ of Rb$Ln_2$Fe$_4$As$_4$O$_2$ as a function of ionic radii of $Ln^{3+}$ ($CN$ = 8). Expectedly, both $a$ and $c$ axes decrease with decreasing the ionic radius. With careful examination, one sees that the cell parameters for $Ln$ = Gd are slightly larger than expected. This might be related to the half filling of 4$f$ level for Gd$^{3+}$.

To investigate the lattice match effect, we also plot the average values of the constituent 122-type and 1111-type unit cells, i.e. $(a_{122} + a_{1111})/2$ and $c_{122} + 2c_{1111}$. Indeed, $a$ and $c$ basically meet the expected values of $(a_{122} + a_{1111})/2$ and $c_{122} + 2c_{1111}$, respectively. One expects that better lattice match would result in a more precise coincidence. However, the best coincidence is seen for $Ln$ = Tb, albeit the lattice match is not the best (see Fig.~\ref{match}). This can be explained by the charge homogenization which leads to an increase in $a_{122}$, and simultaneously, a decrease in $a_{1111}$. That is to say, the case of $Ln$ = Tb actually represents the best lattice match, provided the charge-transfer effect is taken into consideration. As shown in Table~\ref{structure}, indeed, the Fe$-$As1 and Fe$-$As2 bond distances (and other parameters including the As height and As$-$Fe$-$As bong angle) for $Ln$ = Tb are almost identical. Noted also is that the difference, $d_{\mathrm{Fe}-\mathrm{As1}} - d_{\mathrm{Fe}-\mathrm{As2}}$, tends to decrease, and changes its sign, from $Ln$ = Sm to $Ln$ = Ho, which is in accordance with the data crossings at $Ln$ = Tb in Fig.~\ref{lattice}.

\begin{figure}
\center
\includegraphics[width=10cm]{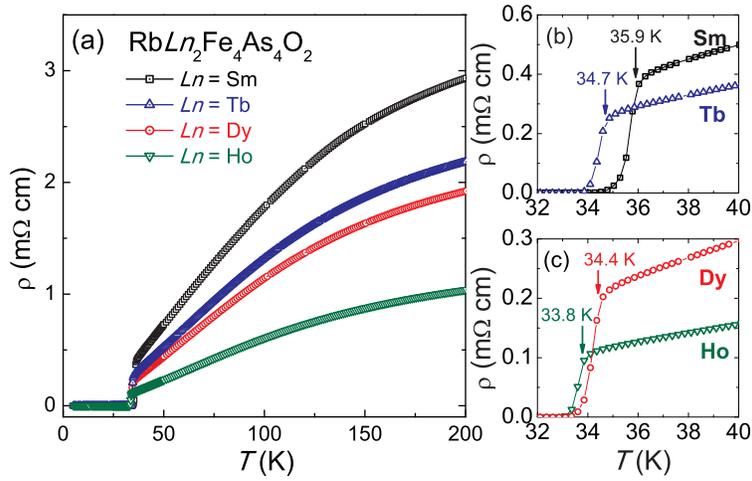}
\caption{\label{rt} (a) Temperature dependence of resistivity for the Rb$Ln$$_2$Fe$_4$As$_4$O$_2$ ($Ln$ = Sm, Tb, Dy and Ho) polycrystalline samples. Superconducting transitions are more clearly displayed in the right-side panels (b) and (c).}
\end{figure}

Figure~\ref{rt} shows resistivity data, $\rho(T)$, of the as-prepared Rb$Ln_2$Fe$_4$As$_4$O$_2$ ($Ln$ = Sm, Tb, Dy and Ho) polycrystals. All the samples show a metallic behavior characterized by a round-shape dependence at around 150 K and a linear relation below $\sim$75 K. The round-shape $\rho(T)$ behavior, which serves as a common characteristic of hole-doped IBS\cite{whh,johrendt,wzc}, is in contrast with the usual linear resistivity arising from electron-phonon scattering. The phenomenon could reflect an incoherent-to-coherent crossover that is in relation with an emergent Kondo-lattice effect\cite{wutao}. The linear $\rho(T)$ behavior below 75 K is also different with those expected for electron-phonon and/or electron-electron scattering. It could represents a possible non-Fermi liquid behavior. The linearity stops when superconductivity sets in at $T_{\mathrm{c}}^{\mathrm{onset}}$ = 33.8 - 35.9 K. The $T_{\mathrm{c}}^{\mathrm{onset}}$ value decreases monotonically from $Ln$ = Sm to Tb, Dy and Ho (the variation of $T_{\mathrm{c}}$ on crystal structure and lanthanide magnetism will be further discussed later on). Coincidently, the room-temperature resistivity and, the resistivity just above $T_{\mathrm{c}}$ in particular, decrease in the same manner. That is to say, $T_{\mathrm{c}}$ and the normal-state resistivity are positively correlated. If the resistivity is significantly contributed from non-phonon scatterings, as argued above, the correlation between $T_{\mathrm{c}}$ and the normal-state resistivity suggests a non-electron-phonon mechanism (such as spin-fluctuation mediated superconductivity) for the occurrence of superconductivity.

\begin{figure}
\center
\includegraphics[width=10cm]{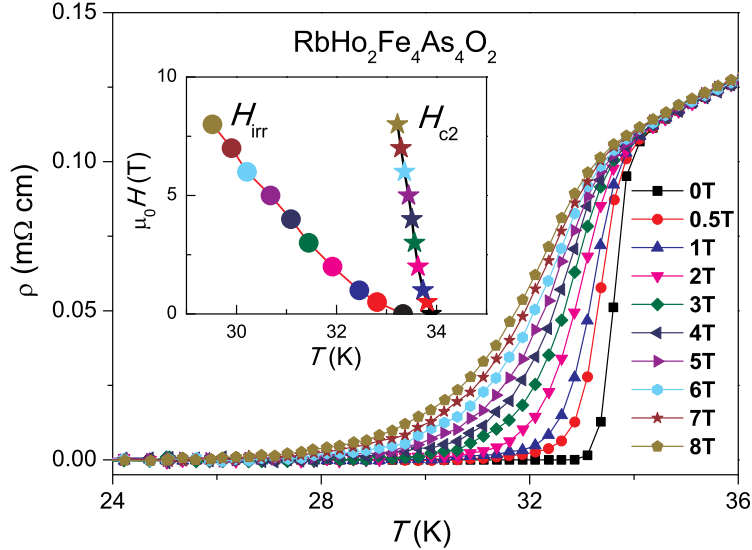}
\caption{\label{mr} Superconducting resistive transitions under magnetic fields for the RbHo$_2$Fe$_4$As$_4$O$_2$ polycrystalline sample. The inset shows the extracted upper critical field $H_{\mathrm{c2}}$ and the irreversible field $H_{\mathrm{irr}}$ as a function of temperature.}
\end{figure}

One of the striking properties in 12442-type superconductors is that the initial slope of upper critical field, $|\mu_0$d$H_{\mathrm{c2}}$/d$T|$, is exceptionally large among IBS\cite{wzc,wzc-scm,wzc-Gd}. For example, the slope value for RbGd$_2$Fe$_4$As$_4$O$_2$ polycrystals achieves 16.5 T/K\cite{wzc-Gd}. To verify the commonality, we measured the magnetoresistance of a representative sample of RbHo$_2$Fe$_4$As$_4$O$_2$. As shown in Fig.~\ref{mr}, the superconducting onset transition shifts very mildly to lower temperatures under external magnetic fields, and simultaneously, the transition becomes significantly broadened with a long tail. To parameterize the field-dependent superconducting transitions, one may extract the upper critical field ($H_{\mathrm{c2}}$) and the irreversible field ($H_{\mathrm{irr}}$). Using the conventional criteria of 90\% and 1\% of the extrapolated normal-state resistivity, the transition temperatures as functions of $H_{\mathrm{c2}}$ and $H_{\mathrm{irr}}$ can be determined. The $H_{\mathrm{c2}}(T)$ and $H_{\mathrm{irr}}(T)$ data thus derived are plotted in the inset of Fig.~\ref{mr}. One sees a steep $H_{\mathrm{c2}}(T)$ line with a slope of 12.5$\pm0.6$ T/K, which is significantly larger than other class of IBS including its relatives, 1144-type CaKFe$_4$As$_4$\cite{canfield1}, RbEuFe$_4$As$_4$\cite{RbEu1144} and CsEuFe$_4$As$_4$\cite{CsEu1144}.

Since the $|\mu_0$d$H_{\mathrm{c2}}$/d$T|$ value is proportional to the orbitally limited upper critical field at zero temperature, $H_{\mathrm{c2}}^{\mathrm{orb}}(0)\approx\frac{\Phi_{0}}{\xi_{i}\xi_{j}}$, where $\Phi_0$ is the magnetic-flux quantum, $\xi_{i}$ and $\xi_{j}$ refer to the coherence lengths perpendicular to the field direction, one may immediately figure out that the coherence length, especially the one along the $c$ axis, should be remarkably smaller than those of other class IBS. The short coherence length is probably originated from the enhanced two dimensionality in relation with the insulating spacer layers. Indeed, the large gap between $H_{\mathrm{c2}}(T)$ and $H_{\mathrm{irr}}(T)$ curves also dictates the weak interlayer coupling related to a short coherence length along the $c$ axis.

\begin{figure*}
  \center
  \includegraphics[width=16cm]{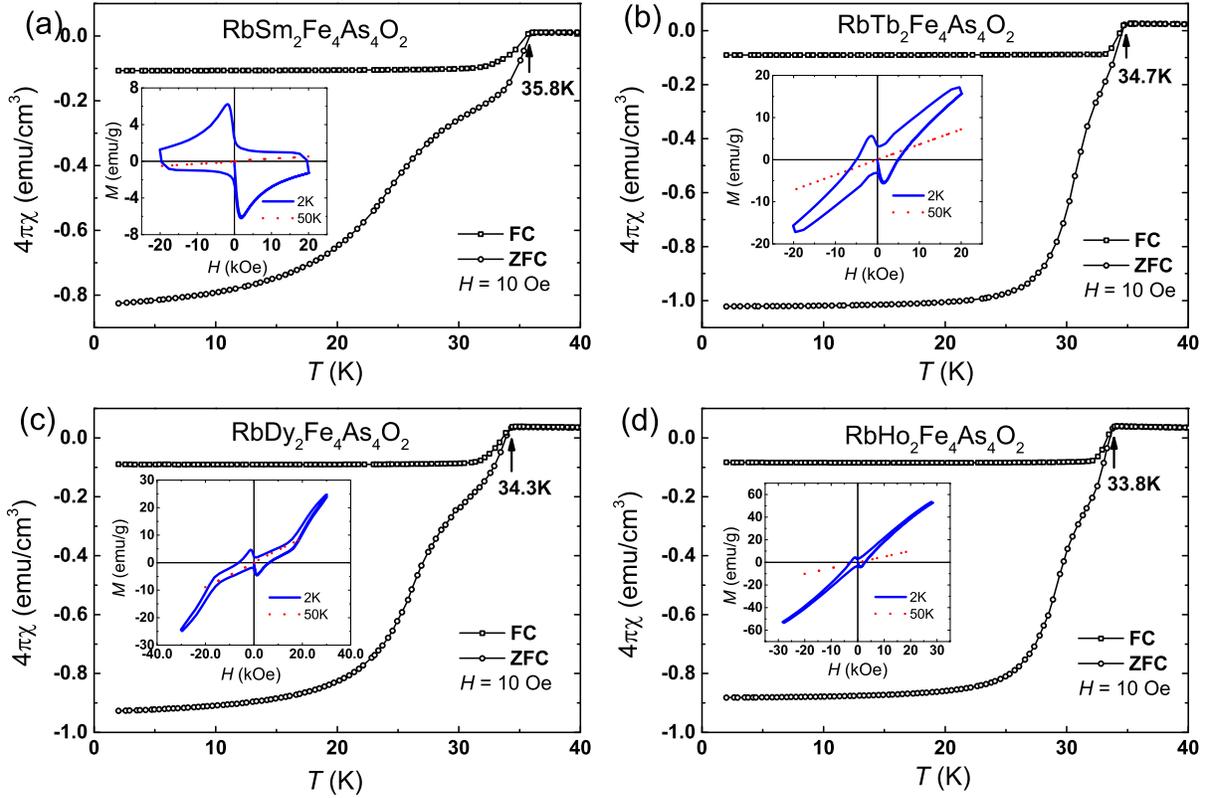}
  \caption{\label{magnetic} Superconductivity in Rb$Ln_2$Fe$_4$As$_4$O$_2$ ($Ln$ = Sm, Tb, Dy and Ho) evidenced by the dc magnetic susceptibility measured at $H$ = 10 Oe in field-cooling (FC) and zero-field-cooling (ZFC) modes. Note that the data were corrected by removing the demagnetization effect. The insets show the isothermal magnetizations at 2 and 50 K.}
\end{figure*}

Bulk superconductivity in Rb$Ln$$_2$Fe$_4$As$_4$O$_2$ is confirmed by the dc magnetic susceptibility shown in Fig.~\ref{magnetic}. Both ZFC and FC data show strong diamagnetism below the superconducting transitions. The onset transition temperatures are 35.8, 34.7, 34.3 and 33.8 K, respectively, for $Ln$ = Sm, Tb, Dy and Ho. The magnetic shielding volume fractions, i.e. 4$\pi\chi$ values in the ZFC mode, are all above 80\% at 2 K. The magnetic repulsion fraction is greatly reduced to about 10\%, which is due to magnetic-flux pinning effect. The flux pinning scenario is further demonstrated by the obvious magnetic hysteresis in the superconducting state (see insets of Fig.~\ref{magnetic}). One also notes that, apparently, there is a step-like anomaly below $T_{\mathrm{c}}$ in the ZFC data, which is absent for the FC data. This phenomenon is ascribed to the effect of intergrain weak links, which often appears for polycrystalline samples of extremely type-II superconductors.

The isothermal magnetization data, shown in the insets of Fig.~\ref{magnetic}, reflect the local-moment magnetism of $Ln^{3+}$ as well. At 50 K (above $T_{\mathrm{c}}$), the $M(H)$ relation is essentially linear with a slope (i.e. magnetic susceptibility) depending on $Ln^{3+}$. The magnetic susceptibility is dominantly contributed from the Curie-Weiss paramagnetism of $Ln^{3+}$ moments. The paramagnetic component is also evident in the superconducting state, as can be seen in the superconducting magnetic hysteresis at 2 K, especially for $Ln$ = Tb, Dy and Ho. In the case of $Ln$ = Dy, the magnetic hysteresis is superposed by a metamagnetic transition at about 20 kOe. Note that the Dy$^{3+}$ magnetic moments in DyFeAsO become antiferromagnetically ordered below $\sim$10 K\cite{lyk-Dy}.

\begin{figure*}
\center
\includegraphics[width=16cm]{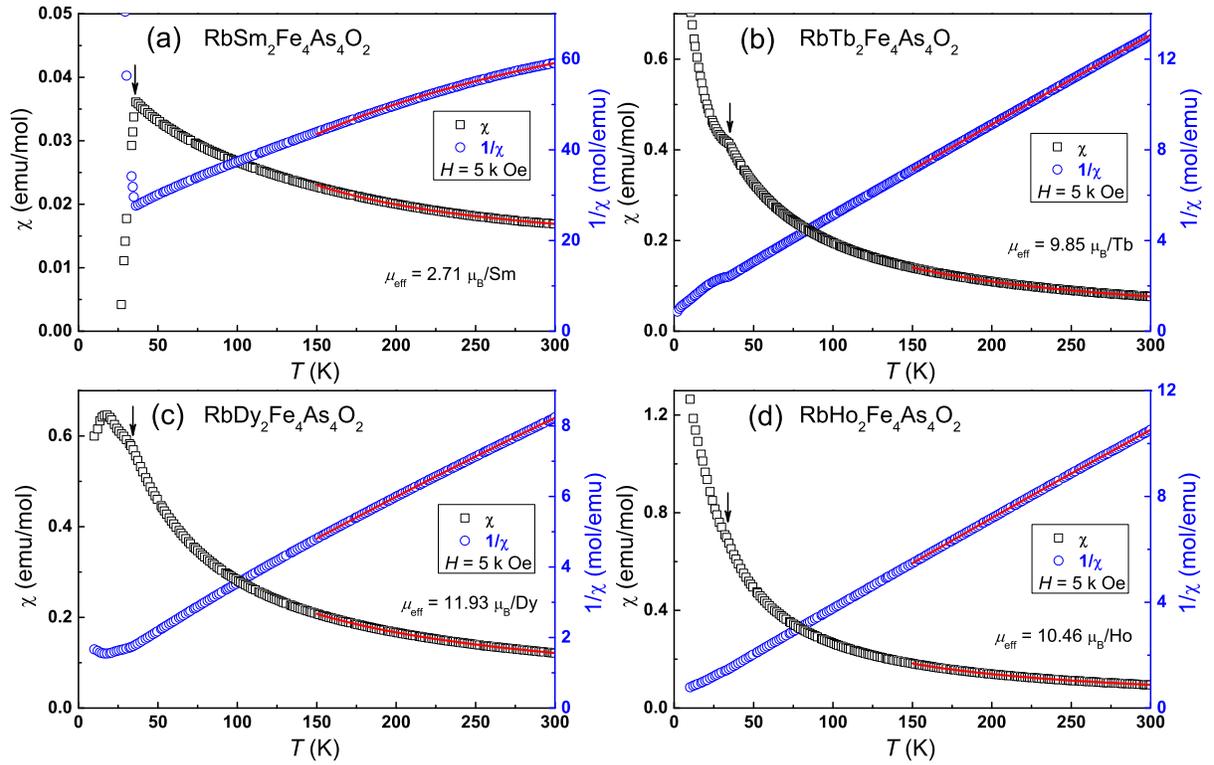}
\caption{\label{chi} Temperature dependence of magnetic susceptibility in Rb$Ln_2$Fe$_4$As$_4$O$_2$ ($Ln$ = Sm, Tb, Dy and Ho). Superconducting transitions are marked by arrows. The right axis is used for showing the reciprocal of the susceptibility. The data between 150 and 300 K are fitted with Curie-Weiss law, from which the effective magnetic moments of $Ln^{3+}$ are obtained.}
\end{figure*}

To further investigate the local-moment magnetism of $Ln^{3+}$, we measured the normal-state magnetic susceptibility of Rb$Ln_2$Fe$_4$As$_4$O$_2$ with an applied field of 5 kOe, as displayed in Figs.~\ref{chi}. The local-moment paramagnetism is confirmed by the linearity in 1/$\chi$. Then, one may be able to extract the effective magnetic moments of $Ln^{3+}$ by a data fitting using the expression, $\chi = \chi_{0}+C/(T-\theta_{\mathrm{p}})$, where $\chi_{0}$ stands for the temperature-independent term, $C$ is the Curie constant, and $\theta_{\mathrm{p}}$ denotes the paramagnetic Curie temperature. To minimize the possible influence from crystal-field effect, we only fit the high-temperature data (150 K $\leq T \leq$ 300 K). The fitting yields effective magnetic moments of 2.71, 9.85, 11.93, 10.46 $\mu_{\mathrm{B}}$/$Ln$-atom for $Ln$ = Sm, Tb, Dy and Ho, respectively. The result basically meets the theoretical value of $g_{J}\sqrt{J(J+1)}$ ($J$ is the quantum number of total angular momentum) for $Ln$ = Tb, Dy and Ho. Note that the discrepancy for $Ln$ = Sm (the experimental value of effective moment is much bigger than the theoretical one) is frequently seen, which is due to low-lying excite states with different $J$ from the ground states\cite{blundell}.


\begin{figure}
\center
\includegraphics[width=10cm]{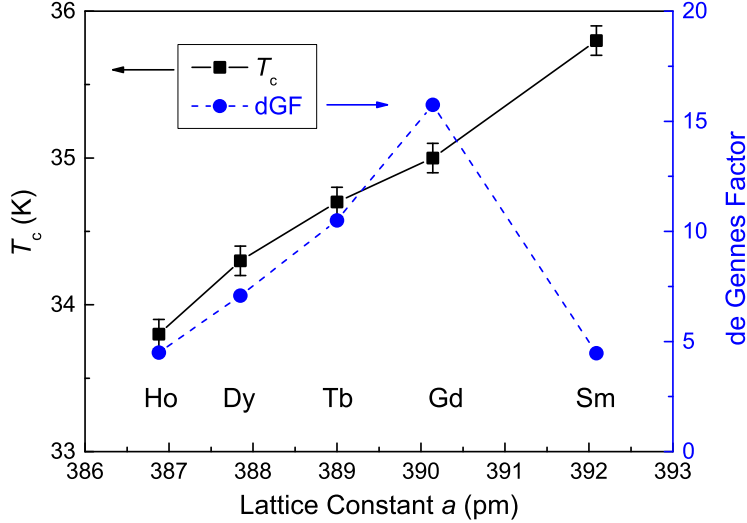}
\caption{\label{tc1} Lattice constant $a$ vs. $T_{\mathrm{c}}$ (left axis) and de Gennes factor (right axis) in Rb$Ln_2$Fe$_4$As$_4$O$_2$ ($Ln$ = Sm, Gd, Tb, Dy and Ho).}
\end{figure}

We found that the $T_{\mathrm{c}}$ value in Rb$Ln_2$Fe$_4$As$_4$O$_2$ remains unchanged (within $\pm$ 0.1 K), irrespective of sample's purity. This fact suggests that Rb$Ln_2$Fe$_4$As$_4$O$_2$ is a line compound which bears the same hole-doping level of 25\%, similar to the case in 1144-type superconductors\cite{1144}. Therefore, it is of meaning to study the possible factors that influence $T_{\mathrm{c}}$. Fig.~\ref{tc1} shows $T_{\mathrm{c}}$ as a function of lattice constant $a$ in Rb$Ln_2$Fe$_4$As$_4$O$_2$. One sees a monotonic increase of $T_{\mathrm{c}}$ with increasing $a$. Notably, however, the $T_{\mathrm{c}}$ value for $Ln$ = Gd is slightly lower than expected from the tendency. This anomaly could be caused by the Gd$^{3+}$ magnetism which exhibits the biggest value of de Gennes factor, $(g_{J}-1)^{2}J(J+1)$, as shown on the right axis in Fig.~\ref{tc1}. The de Gennes factor measures the magnetic pair-breaking strength. Hence the slight decrease in $T_{\mathrm{c}}$ for $Ln$ = Gd actually dictates that $Ln^{3+}$ magnetism hardly influence the $T_{\mathrm{c}}$ in Rb$Ln_2$Fe$_4$As$_4$O$_2$.

\begin{figure*}
\center
\includegraphics[width=16cm]{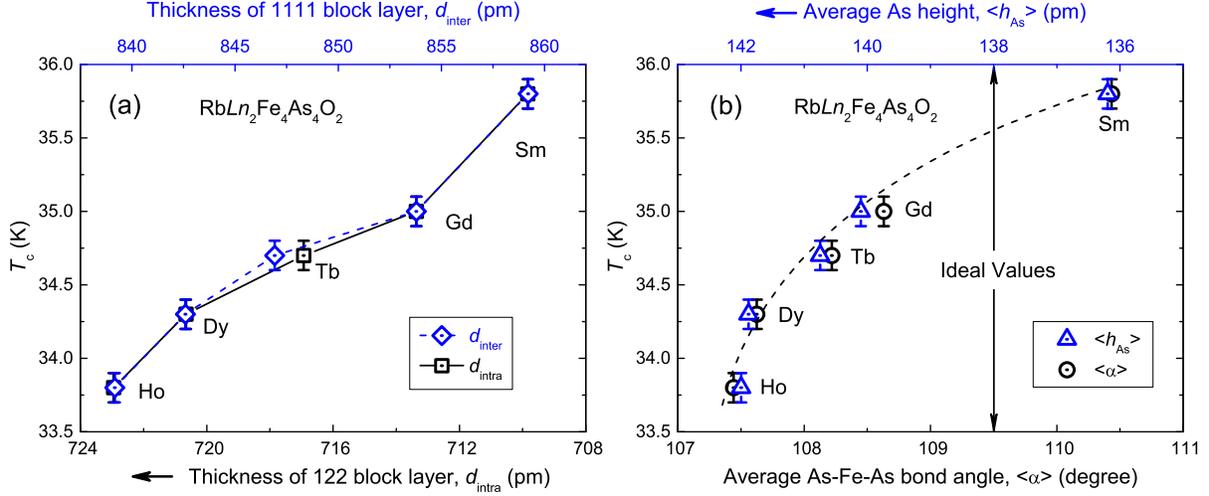}
\caption{\label{tc2} (a) $T_{\mathrm{c}}$ vs. thicknesses of the 122- and 1111-like blocks in Rb$Ln_2$Fe$_4$As$_4$O$_2$ ($Ln$ = Sm, Gd, Tb, Dy and Ho). (b) Influence of the average As$-$Fe$-$As bond angle and the As height from the Fe plane on $T_{\mathrm{c}}$. The dashed line is a guide to the eye. The vertical line with arrows represents the values that are assumed to give the maximum $T_{\mathrm{c}}$.}
\end{figure*}

The lattice-size dependence of $T_{\mathrm{c}}$ above contradicts with the case in $Ak$Ca$_2$Fe$_4$As$_4$F$_2$ ($Ak$ = K, Rb and Cs)\cite{wzc-scm}. Therefore, lattice constants are not good parameters that control $T_{\mathrm{c}}$. In the $Ak$Ca$_2$Fe$_4$As$_4$F$_2$ series, we found that the spacings of Fe$_2$As$_2$ layers seem to be relevant: $T_{\mathrm{c}}$ increases with the decrease (increase) of intra(inter)-bilayer spacing, $d_{\mathrm{intra}}$ ($d_{\mathrm{inter}}$). Note that $d_{\mathrm{intra}}$ and $d_{\mathrm{inter}}$ also measure the thickness of the 122- and 1111-like blocks, respectively (see Fig.~\ref{match}). For Rb$Ln_2$Fe$_4$As$_4$O$_2$, a similar relation appears, as shown in Fig.~\ref{tc2}(a). The slight deviation for $Ln$ = Gd could be due to the large de Gennes factor of Gd$^{3+}$ as mentioned above. The observation of relationship between Fe$_2$As$_2$-layer spacings and $T_{\mathrm{c}}$ suggests the role of interlayer coupling on superconductivity.

As far as a single Fe$_2$As$_2$ layer is concerned, in fact, the structural correlations of $T_{\mathrm{c}}$ are widely discussed in terms of the As$-$Fe$-$As bond angle, $\alpha$, and/or the As height from the Fe plane, $h_{\mathrm{As}}$\cite{zhaoj,lee,mizuguchi}. It is concluded that the maximum $T_{\mathrm{c}}$ appears at $\alpha$ = 109.5$^{\circ}$ or $h_{\mathrm{As}}$ = 138 pm. As for Rb$Ln_2$Fe$_4$As$_4$O$_2$, we have two distinct As sites, which give two values for each parameter. It turns out that the difference in $\alpha$ or $h_{\mathrm{As}}$ does not correlate with $T_{\mathrm{c}}$. We thus consider the average values, $<\alpha>$  and $<h_{\mathrm{As}}>$ (this is reasonable because there is only one Fe site). Strikingly, a monotonic relation is found for both $<\alpha>$ and $<h_{\mathrm{As}}>$, as shown in Fig.~\ref{tc2}(b). No signature of $T_{\mathrm{c}}$ optimization is evident at $\alpha$ = 109.5$^{\circ}$ or $h_{\mathrm{As}}$ = 138 pm. Invalidation of the correlations between $T_{\mathrm{c}}$ and the geometry of single Fe$_2$As$_2$ layer is also seen in $Ak$Ca$_2$Fe$_4$As$_4$F$_2$ system\cite{wzc-scm}, which suggests that Fe$_2$As$_2$-layer spacings could be another structural parameter controlling $T_{\mathrm{c}}$.

\section{CONCLUDING REMARKS}

To summarize, we were able to synthesize the quinary Rb$Ln_2$Fe$_4$As$_4$O$_2$ series at ambient pressure for $Ln$ = Sm, Tb, Dy and Ho. The results indicate that lattice match between RbFe$_2$As$_2$ and $Ln$FeAsO, which is modified by the charge homogenization, is crucial for the phase stabilization. In addition, the intergrowth constituents (such as $Ln$FeAsO) themselves should preferably be stable. In this sense, formation of RbHo$_2$Fe$_4$As$_4$O$_2$ is remarkable because HoFeAsO cannot be synthesized in the stoichiometric composition at atmospheric pressure. According to the lattice-match viewpoint, we prospect that Rb$At_2$Fe$_4$As$_4$O$_2$ ($At$ = Np and Pu) and $Ak$$Ln_2$Fe$_4$As$_4$O$_2$ ($Ak$ = K and Cs) are likely to be synthesized for the future.

Like their sister materials, Rb$Ln_2$Fe$_4$As$_4$O$_2$ are featured by double asymmetric Fe$_2$As$_2$ layers that are intrinsically hole doped (0.25 holes/Fe-atom). Bulk superconductivity, instead of SDW order, appears in all the stoichiometric quinary compounds. The $T_\mathrm{c}$ values (from 33.8 to 35.8 K) are higher than those of 1111-type hole-doped superconductors which contain single separate single Fe$_2$As$_2$ layer, yet they are still lower than that of (Ba,K)Fe$_2$As$_2$ which contains infinite Fe$_2$As$_2$ layer. The widely accepted structural parameters related to $T_{\mathrm{c}}$, i.e. As$-$Fe$-$As bond angle and As height from Fe plane, cannot account for the $T_{\mathrm{c}}$ variation. Instead, the Fe$_2$As$_2$-layer spacing seems to be an important factor controlling $T_{\mathrm{c}}$ in 12442 systems. This suggests that interlayer couplings may play an additional role for optimization of superconductivity in IBS.

\begin{acknowledgement}
This work was supported by the National Science Foundation of China (Nos. 11474252 and 11190023) and the National Key Research and Development Program of China (No. 2016YFA0300202).
\end{acknowledgement}

\begin{suppinfo}
CIF files of the crystallographic data of Rb$Ln_2$Fe$_4$As$_4$O$_2$ ($Ln$ = Sm, Gd, Tb, Dy and Ho).

\end{suppinfo}


\bibliography{12442_cm}

\providecommand*\mcitethebibliography{\thebibliography}
\csname @ifundefined\endcsname{endmcitethebibliography}
  {\let\endmcitethebibliography\endthebibliography}{}
\begin{mcitethebibliography}{34}
\providecommand*\natexlab[1]{#1}
\providecommand*\mciteSetBstSublistMode[1]{}
\providecommand*\mciteSetBstMaxWidthForm[2]{}
\providecommand*\mciteBstWouldAddEndPuncttrue
  {\def\EndOfBibitem{\unskip.}}
\providecommand*\mciteBstWouldAddEndPunctfalse
  {\let\EndOfBibitem\relax}
\providecommand*\mciteSetBstMidEndSepPunct[3]{}
\providecommand*\mciteSetBstSublistLabelBeginEnd[3]{}
\providecommand*\EndOfBibitem{}
\mciteSetBstSublistMode{f}
\mciteSetBstMaxWidthForm{subitem}{(\alph{mcitesubitemcount})}
\mciteSetBstSublistLabelBeginEnd
  {\mcitemaxwidthsubitemform\space}
  {\relax}
  {\relax}

\bibitem[Hosono and Kuroki(2015)Hosono, and Kuroki]{hosono-pc}
Hosono,~H.; Kuroki,~K. \emph{Phys. C} \textbf{2015}, \emph{514}, 399--422\relax
\mciteBstWouldAddEndPuncttrue
\mciteSetBstMidEndSepPunct{\mcitedefaultmidpunct}
{\mcitedefaultendpunct}{\mcitedefaultseppunct}\relax
\EndOfBibitem
\bibitem[Luo and Chen(2015)Luo, and Chen]{cxh-scm}
Luo,~X.; Chen,~X. \emph{Sci. China Mater.} \textbf{2015}, \emph{58},
  77--89\relax
\mciteBstWouldAddEndPuncttrue
\mciteSetBstMidEndSepPunct{\mcitedefaultmidpunct}
{\mcitedefaultendpunct}{\mcitedefaultseppunct}\relax
\EndOfBibitem
\bibitem[Jiang et~al.(2013)Jiang, Sun, Xu, and Cao]{jh}
Jiang,~H.; Sun,~Y.-L.; Xu,~Z.-A.; Cao,~G.-H. \emph{Chin. Phys. B}
  \textbf{2013}, \emph{22}, 087410\relax
\mciteBstWouldAddEndPuncttrue
\mciteSetBstMidEndSepPunct{\mcitedefaultmidpunct}
{\mcitedefaultendpunct}{\mcitedefaultseppunct}\relax
\EndOfBibitem
\bibitem[Kamihara et~al.(2008)Kamihara, Watanabe, Hirano, and Hosono]{hosono}
Kamihara,~Y.; Watanabe,~T.; Hirano,~M.; Hosono,~H. \emph{J. Am. Chem. Soc.}
  \textbf{2008}, \emph{130}, 3296--7297\relax
\mciteBstWouldAddEndPuncttrue
\mciteSetBstMidEndSepPunct{\mcitedefaultmidpunct}
{\mcitedefaultendpunct}{\mcitedefaultseppunct}\relax
\EndOfBibitem
\bibitem[Wen et~al.(2008)Wen, Mu, Fang, Yang, and Zhu]{whh}
Wen,~H.-H.; Mu,~G.; Fang,~L.; Yang,~H.; Zhu,~X. \emph{EPL} \textbf{2008},
  \emph{82}, 17009\relax
\mciteBstWouldAddEndPuncttrue
\mciteSetBstMidEndSepPunct{\mcitedefaultmidpunct}
{\mcitedefaultendpunct}{\mcitedefaultseppunct}\relax
\EndOfBibitem
\bibitem[Ren et~al.(2009)Ren, Tao, Jiang, Feng, Wang, Dai, Cao, and Xu]{rz}
Ren,~Z.; Tao,~Q.; Jiang,~S.; Feng,~C.; Wang,~C.; Dai,~J.; Cao,~G.; Xu,~Z.
  \emph{Phys. Rev. Lett.} \textbf{2009}, \emph{102}, 137002\relax
\mciteBstWouldAddEndPuncttrue
\mciteSetBstMidEndSepPunct{\mcitedefaultmidpunct}
{\mcitedefaultendpunct}{\mcitedefaultseppunct}\relax
\EndOfBibitem
\bibitem[Cao et~al.(2010)Cao, Ma, Wang, Sun, Bao, Jiang, Luo, Feng, Zhou, Xie,
  Hu, Wei, Nowik, Felner, Zhang, Xu, and Zhang]{cgh}
Cao,~G.-H. et~al.  \emph{Phys. Rev. B} \textbf{2010}, \emph{82}, 104518\relax
\mciteBstWouldAddEndPuncttrue
\mciteSetBstMidEndSepPunct{\mcitedefaultmidpunct}
{\mcitedefaultendpunct}{\mcitedefaultseppunct}\relax
\EndOfBibitem
\bibitem[Sun et~al.(2012)Sun, Jiang, Zhai, Bao, Jiao, Tao, Shen, Zeng, Xu, and
  Cao]{syl}
Sun,~Y.~L.; Jiang,~H.; Zhai,~H.~F.; Bao,~J.~K.; Jiao,~W.~H.; Tao,~Q.;
  Shen,~C.~Y.; Zeng,~Y.~W.; Xu,~Z.~A.; Cao,~G.~H. \emph{J. Am. Chem. Soc.}
  \textbf{2012}, \emph{134}, 12893--12896\relax
\mciteBstWouldAddEndPuncttrue
\mciteSetBstMidEndSepPunct{\mcitedefaultmidpunct}
{\mcitedefaultendpunct}{\mcitedefaultseppunct}\relax
\EndOfBibitem
\bibitem[Iyo et~al.(2016)Iyo, Kawashima, Kinjo, Nishio, Ishida, Fujihisa,
  Gotoh, Kihou, Eisaki, and Yoshida]{1144}
Iyo,~A.; Kawashima,~K.; Kinjo,~T.; Nishio,~T.; Ishida,~S.; Fujihisa,~H.;
  Gotoh,~Y.; Kihou,~K.; Eisaki,~H.; Yoshida,~Y. \emph{J. Am. Chem. Soc.}
  \textbf{2016}, \emph{138}, 3410--3415\relax
\mciteBstWouldAddEndPuncttrue
\mciteSetBstMidEndSepPunct{\mcitedefaultmidpunct}
{\mcitedefaultendpunct}{\mcitedefaultseppunct}\relax
\EndOfBibitem
\bibitem[Kawashima et~al.(2016)Kawashima, Kinjo, Nishio, Ishida, Fujihisa,
  Gotoh, Kihou, Eisaki, Yoshida, and Iyo]{Eu1144}
Kawashima,~K.; Kinjo,~T.; Nishio,~T.; Ishida,~S.; Fujihisa,~H.; Gotoh,~Y.;
  Kihou,~K.; Eisaki,~H.; Yoshida,~Y.; Iyo,~A. \emph{J. Phys. Soc. Jpn.}
  \textbf{2016}, \emph{85}, 064710\relax
\mciteBstWouldAddEndPuncttrue
\mciteSetBstMidEndSepPunct{\mcitedefaultmidpunct}
{\mcitedefaultendpunct}{\mcitedefaultseppunct}\relax
\EndOfBibitem
\bibitem[Liu et~al.(2016)Liu, Liu, Tang, Jiang, Wang, Ablimit, Jiao, Tao, Feng,
  Xu, and Cao]{RbEu1144}
Liu,~Y.; Liu,~Y.-B.; Tang,~Z.-T.; Jiang,~H.; Wang,~Z.-C.; Ablimit,~A.;
  Jiao,~W.-H.; Tao,~Q.; Feng,~C.-M.; Xu,~Z.-A.; Cao,~G.-H. \emph{Phys. Rev. B}
  \textbf{2016}, \emph{93}, 214503\relax
\mciteBstWouldAddEndPuncttrue
\mciteSetBstMidEndSepPunct{\mcitedefaultmidpunct}
{\mcitedefaultendpunct}{\mcitedefaultseppunct}\relax
\EndOfBibitem
\bibitem[Liu et~al.(2016)Liu, Liu, Chen, Tang, Jiao, Tao, Xu, and
  Cao]{CsEu1144}
Liu,~Y.; Liu,~Y.-B.; Chen,~Q.; Tang,~Z.-T.; Jiao,~W.-H.; Tao,~Q.; Xu,~Z.-A.;
  Cao,~G.-H. \emph{Sci. Bull.} \textbf{2016}, \emph{61}, 1213--1220\relax
\mciteBstWouldAddEndPuncttrue
\mciteSetBstMidEndSepPunct{\mcitedefaultmidpunct}
{\mcitedefaultendpunct}{\mcitedefaultseppunct}\relax
\EndOfBibitem
\bibitem[Wang et~al.(2016)Wang, He, Wu, Tang, Liu, Ablimit, Feng, and Cao]{wzc}
Wang,~Z.-C.; He,~C.-Y.; Wu,~S.-Q.; Tang,~Z.-T.; Liu,~Y.; Ablimit,~A.;
  Feng,~C.-M.; Cao,~G.-H. \emph{J. Am. Chem. Soc.} \textbf{2016}, \emph{138},
  7856--7859\relax
\mciteBstWouldAddEndPuncttrue
\mciteSetBstMidEndSepPunct{\mcitedefaultmidpunct}
{\mcitedefaultendpunct}{\mcitedefaultseppunct}\relax
\EndOfBibitem
\bibitem[Wang et~al.(2017)Wang, He, Tang, Wu, and Cao]{wzc-scm}
Wang,~Z.-C.; He,~C.-Y.; Tang,~Z.-T.; Wu,~S.-Q.; Cao,~G.-H. \emph{Sci. China
  Mater.} \textbf{2017}, \emph{60}, 83--89\relax
\mciteBstWouldAddEndPuncttrue
\mciteSetBstMidEndSepPunct{\mcitedefaultmidpunct}
{\mcitedefaultendpunct}{\mcitedefaultseppunct}\relax
\EndOfBibitem
\bibitem[Wang et~al.(2017)Wang, He, Wu, Tang, Liu, Ablimit, Tao, Feng, Xu, and
  Cao]{wzc-Gd}
Wang,~Z.-C.; He,~C.-Y.; Wu,~S.-Q.; Tang,~Z.-T.; Liu,~Y.; Ablimit,~A.; Tao,~Q.;
  Feng,~C.-M.; Xu,~Z.-A.; Cao,~G.-H. \emph{J. Phys.: Condens. Matt.}
  \textbf{2017}, \emph{29}, 11LT01\relax
\mciteBstWouldAddEndPuncttrue
\mciteSetBstMidEndSepPunct{\mcitedefaultmidpunct}
{\mcitedefaultendpunct}{\mcitedefaultseppunct}\relax
\EndOfBibitem
\bibitem[Nitsche et~al.(2010)Nitsche, Jesche, Hieckmann, Doert, and Ruck]{LnO}
Nitsche,~F.; Jesche,~A.; Hieckmann,~E.; Doert,~T.; Ruck,~M. \emph{Phys. Rev. B}
  \textbf{2010}, \emph{82}, 134514\relax
\mciteBstWouldAddEndPuncttrue
\mciteSetBstMidEndSepPunct{\mcitedefaultmidpunct}
{\mcitedefaultendpunct}{\mcitedefaultseppunct}\relax
\EndOfBibitem
\bibitem[Muraba et~al.(2014)Muraba, Matsuishi, and Hosono]{CaH}
Muraba,~Y.; Matsuishi,~S.; Hosono,~H. \emph{J. Phys. Soc. Jpn.} \textbf{2014},
  \emph{83}, 033705\relax
\mciteBstWouldAddEndPuncttrue
\mciteSetBstMidEndSepPunct{\mcitedefaultmidpunct}
{\mcitedefaultendpunct}{\mcitedefaultseppunct}\relax
\EndOfBibitem
\bibitem[Matsuishi et~al.(2008)Matsuishi, Inoue, Nomura, Yanagi, Hirano, and
  Hosono]{CaF}
Matsuishi,~S.; Inoue,~Y.; Nomura,~T.; Yanagi,~H.; Hirano,~M.; Hosono,~H.
  \emph{J. Am. Chem. Soc.} \textbf{2008}, \emph{130}, 14428--14429\relax
\mciteBstWouldAddEndPuncttrue
\mciteSetBstMidEndSepPunct{\mcitedefaultmidpunct}
{\mcitedefaultendpunct}{\mcitedefaultseppunct}\relax
\EndOfBibitem
\bibitem[Zhu et~al.(2009)Zhu, Han, Cheng, Mu, Shen, Zeng, and Wen]{SrF}
Zhu,~X.~Y.; Han,~F.; Cheng,~P.; Mu,~G.; Shen,~B.; Zeng,~B.; Wen,~H.~H.
  \emph{Phys. C} \textbf{2009}, \emph{469}, 381--384\relax
\mciteBstWouldAddEndPuncttrue
\mciteSetBstMidEndSepPunct{\mcitedefaultmidpunct}
{\mcitedefaultendpunct}{\mcitedefaultseppunct}\relax
\EndOfBibitem
\bibitem[Klimczuk et~al.(2012)Klimczuk, Walker, Springell, Shick, Hill,
  Gaczy\ifmmode~\acute{n}\else \'{n}\fi{}ski, Gofryk, Kimber, Ritter, Colineau,
  Griveau, Bou\"exi\`ere, Eloirdi, Cava, and Caciuffo]{NpO}
Klimczuk,~T.; Walker,~H.~C.; Springell,~R.; Shick,~A.~B.; Hill,~A.~H.;
  Gaczy\ifmmode~\acute{n}\else \'{n}\fi{}ski,~P.; Gofryk,~K.; Kimber,~S. A.~J.;
  Ritter,~C.; Colineau,~E.; Griveau,~J.-C.; Bou\"exi\`ere,~D.; Eloirdi,~R.;
  Cava,~R.~J.; Caciuffo,~R. \emph{Phys. Rev. B} \textbf{2012}, \emph{85},
  174506\relax
\mciteBstWouldAddEndPuncttrue
\mciteSetBstMidEndSepPunct{\mcitedefaultmidpunct}
{\mcitedefaultendpunct}{\mcitedefaultseppunct}\relax
\EndOfBibitem
\bibitem[Klimczuk et~al.(2012)Klimczuk, Shick, Springell, Walker, Hill,
  Colineau, Griveau, Bou\"exi\`ere, Eloirdi, and Caciuffo]{PuO}
Klimczuk,~T.; Shick,~A.~B.; Springell,~R.; Walker,~H.~C.; Hill,~A.~H.;
  Colineau,~E.; Griveau,~J.-C.; Bou\"exi\`ere,~D.; Eloirdi,~R.; Caciuffo,~R.
  \emph{Phys. Rev. B} \textbf{2012}, \emph{86}, 174510\relax
\mciteBstWouldAddEndPuncttrue
\mciteSetBstMidEndSepPunct{\mcitedefaultmidpunct}
{\mcitedefaultendpunct}{\mcitedefaultseppunct}\relax
\EndOfBibitem
\bibitem[Wang et~al.(2016)Wang, Wang, Mei, Li, Li, Tang, Liu, Zhang, Zhai, Xu,
  and Cao]{wc2016}
Wang,~C.; Wang,~Z.-C.; Mei,~Y.-X.; Li,~Y.-K.; Li,~L.; Tang,~Z.-T.; Liu,~Y.;
  Zhang,~P.; Zhai,~H.-F.; Xu,~Z.-A.; Cao,~G.-H. \emph{J. Am. Chem. Soc.}
  \textbf{2016}, \emph{138}, 2170--2173\relax
\mciteBstWouldAddEndPuncttrue
\mciteSetBstMidEndSepPunct{\mcitedefaultmidpunct}
{\mcitedefaultendpunct}{\mcitedefaultseppunct}\relax
\EndOfBibitem
\bibitem[Shannon(1976)]{shannon}
Shannon,~R.~D. \emph{Acta Crystallogr. Sect. A} \textbf{1976}, \emph{32},
  751--767\relax
\mciteBstWouldAddEndPuncttrue
\mciteSetBstMidEndSepPunct{\mcitedefaultmidpunct}
{\mcitedefaultendpunct}{\mcitedefaultseppunct}\relax
\EndOfBibitem
\bibitem[Izumi and Momma(2007)Izumi, and Momma]{rietan}
Izumi,~F.; Momma,~K. \emph{Applied Crystallography XX}; Solid State Phenomena;
  Elsevier Ltd., 2007; Chapter 130, pp 15--20\relax
\mciteBstWouldAddEndPuncttrue
\mciteSetBstMidEndSepPunct{\mcitedefaultmidpunct}
{\mcitedefaultendpunct}{\mcitedefaultseppunct}\relax
\EndOfBibitem
\bibitem[Walters et~al.(2015)Walters, Walker, Springell, Krisch, Bosak, Hill,
  Zvoriste-Walters, Colineau, Griveau, Bouexiere, Eloirdi, Caciuffo, and
  Klimczuk]{Np1111}
Walters,~A.~C.; Walker,~H.~C.; Springell,~R.; Krisch,~M.; Bosak,~A.;
  Hill,~A.~H.; Zvoriste-Walters,~C.~E.; Colineau,~E.; Griveau,~J.-C.;
  Bouexiere,~D.; Eloirdi,~R.; Caciuffo,~R.; Klimczuk,~T. \emph{J. Phys.:
  Condens. Matt.} \textbf{2015}, \emph{27}, 325702\relax
\mciteBstWouldAddEndPuncttrue
\mciteSetBstMidEndSepPunct{\mcitedefaultmidpunct}
{\mcitedefaultendpunct}{\mcitedefaultseppunct}\relax
\EndOfBibitem
\bibitem[Rotter et~al.(2008)Rotter, Tegel, and Johrendt]{johrendt}
Rotter,~M.; Tegel,~M.; Johrendt,~D. \emph{Phys. Rev. Lett.} \textbf{2008},
  \emph{101}, 107006\relax
\mciteBstWouldAddEndPuncttrue
\mciteSetBstMidEndSepPunct{\mcitedefaultmidpunct}
{\mcitedefaultendpunct}{\mcitedefaultseppunct}\relax
\EndOfBibitem
\bibitem[Wu et~al.(2016)Wu, Zhao, Wang, Wang, Xiang, Luo, Wu, and Chen]{wutao}
Wu,~Y.~P.; Zhao,~D.; Wang,~A.~F.; Wang,~N.~Z.; Xiang,~Z.~J.; Luo,~X.~G.;
  Wu,~T.; Chen,~X.~H. \emph{Phys. Rev. Lett.} \textbf{2016}, \emph{116},
  147001\relax
\mciteBstWouldAddEndPuncttrue
\mciteSetBstMidEndSepPunct{\mcitedefaultmidpunct}
{\mcitedefaultendpunct}{\mcitedefaultseppunct}\relax
\EndOfBibitem
\bibitem[Meier et~al.(2016)Meier, Kong, Kaluarachchi, Taufour, Jo, Drachuck,
  B\"ohmer, Saunders, Sapkota, Kreyssig, Tanatar, Prozorov, Goldman, Balakirev,
  Gurevich, Bud'ko, and Canfield]{canfield1}
Meier,~W.~R. et~al.  \emph{Phys. Rev. B} \textbf{2016}, \emph{94}, 064501\relax
\mciteBstWouldAddEndPuncttrue
\mciteSetBstMidEndSepPunct{\mcitedefaultmidpunct}
{\mcitedefaultendpunct}{\mcitedefaultseppunct}\relax
\EndOfBibitem
\bibitem[Luo et~al.(2012)Luo, Lin, Li, Tao, Li, Zhu, Cao, and Xu]{lyk-Dy}
Luo,~Y.; Lin,~X.; Li,~Y.; Tao,~Q.; Li,~L.; Zhu,~Z.; Cao,~G.; Xu,~Z.
  \emph{Inter. J. Mod. Phys. B} \textbf{2012}, \emph{26}, 1250207\relax
\mciteBstWouldAddEndPuncttrue
\mciteSetBstMidEndSepPunct{\mcitedefaultmidpunct}
{\mcitedefaultendpunct}{\mcitedefaultseppunct}\relax
\EndOfBibitem
\bibitem[Blundell(2001)]{blundell}
Blundell,~S. \emph{Magnetism in Condensed Matter}; Oxford Unversity Press,
  Oxford, 2001; p~35\relax
\mciteBstWouldAddEndPuncttrue
\mciteSetBstMidEndSepPunct{\mcitedefaultmidpunct}
{\mcitedefaultendpunct}{\mcitedefaultseppunct}\relax
\EndOfBibitem
\bibitem[Zhao et~al.(2008)Zhao, Huang, de~la Cruz, Li, Lynn, Chen, Green, Chen,
  Li, Li, Luo, Wang, and Dai]{zhaoj}
Zhao,~J.; Huang,~Q.; de~la Cruz,~C.; Li,~S.; Lynn,~J.~W.; Chen,~Y.;
  Green,~M.~A.; Chen,~G.~F.; Li,~G.; Li,~Z.; Luo,~J.~L.; Wang,~N.~L.; Dai,~P.
  \emph{Nat. Mater.} \textbf{2008}, \emph{7}, 953--959\relax
\mciteBstWouldAddEndPuncttrue
\mciteSetBstMidEndSepPunct{\mcitedefaultmidpunct}
{\mcitedefaultendpunct}{\mcitedefaultseppunct}\relax
\EndOfBibitem
\bibitem[Lee et~al.(2008)Lee, Iyo, Eisaki, Kito, Fernandez-Diaz, Ito, Kihou,
  Matsuhata, Braden, and Yamada]{lee}
Lee,~C.-H.; Iyo,~A.; Eisaki,~H.; Kito,~H.; Fernandez-Diaz,~M.~T.; Ito,~T.;
  Kihou,~K.; Matsuhata,~H.; Braden,~M.; Yamada,~K. \emph{J. Phys. Soc. Jpn.}
  \textbf{2008}, \emph{77}, 083704\relax
\mciteBstWouldAddEndPuncttrue
\mciteSetBstMidEndSepPunct{\mcitedefaultmidpunct}
{\mcitedefaultendpunct}{\mcitedefaultseppunct}\relax
\EndOfBibitem
\bibitem[Mizuguchi et~al.(2010)Mizuguchi, Hara, Deguchi, Tsuda, Yamaguchi,
  Takeda, Kotegawa, Tou, and Takano]{mizuguchi}
Mizuguchi,~Y.; Hara,~Y.; Deguchi,~K.; Tsuda,~S.; Yamaguchi,~T.; Takeda,~K.;
  Kotegawa,~H.; Tou,~H.; Takano,~Y. \emph{Supercond. Sci. Technol.}
  \textbf{2010}, \emph{23}, 054013\relax
\mciteBstWouldAddEndPuncttrue
\mciteSetBstMidEndSepPunct{\mcitedefaultmidpunct}
{\mcitedefaultendpunct}{\mcitedefaultseppunct}\relax
\EndOfBibitem
\end{mcitethebibliography}

\end{document}